\documentclass{PoS}

\usepackage{enumerate}
\usepackage{amsmath,amssymb}
\usepackage{mathrsfs}
\usepackage{bm}

\title{Confinement/deconfinement transition temperature from the Polyakov loop potential and gauge-invariant gluon mass }

\ShortTitle{Confinement/deconfinement transition}

\author{\speaker{Kei-Ichi Kondo}
\\
        Department of Physics,  
Graduate School of Science, 
Chiba University, Chiba 263-8522, Japan\\
        E-mail: \email{kondok@faculty.chiba-u.jp}}
        
\author{Akihiro Shibata
\\
Computing Research Center, High Energy Accelerator Research Organization (KEK), Tsukuba 305-0801, Japan
}

\abstract{
We give an analytical derivation of the confinement/deconfinement phase transition at finite temperature in the $SU(N)$ Yang-Mills theory in the $D$-dimensional space time for $D>2$. 
For this purpose, we use a novel reformulation of the Yang-Mills theory which allows the gauge-invariant gluonic mass term, 
and calculate analytically the effective potential of the Polyakov loop average concretely for the $SU(2)$ and $SU(3)$ Yang-Mills theories by including the gauge-invariant dynamical gluonic mass $M$.  
For $D=4$, we give an  estimate on the transition temperature $T_d$ as  the ratio $T_d/M$ to the  mass $M$ which has been measured on the lattice at zero temperature and is  calculable also at finite temperature. 
We show that the order of the phase transition at $T_d$ is the second  order for $SU(2)$ and weakly first order for $SU(3)$ Yang-Mills theory.
We elucidate what is the mechanism for quark confinement and deconfinement at finite temperature and why the phase transition  occurs at a certain temperature.
These initial results are obtained easily based on the analytical calculations of the ``one-loop type'' in the first approximation. 
We discuss also how these results are improved to eliminate the artifacts obtained for some thermodynamic observables

}

\FullConference{The 33rd International Symposium on Lattice Field Theory\\
		14 -18 July 2015\\
		Kobe International Conference Center, Kobe, Japan*}

\begin{document}

\section{Introduction}

Quark confinement and chiral-symmetry breaking  are the main subjects to be investigated for understanding the various phases in the gauge theory for strong interactions, namely QCD at finite temperature and density. 
In a previous paper \cite{Kondo10}, we have proposed a theoretical framework to obtain a low-energy effective theory of QCD towards a first-principle derivation of confinement/deconfinement and chiral-symmetry breaking/restoration crossover transitions at finite temperature.
The basic ingredients are a novel reformulation \cite{KKSS15} of Yang-Mills theory and QCD based on new variables 
and the flow equation of the Wetterich type in the framework of the functional renormalization group (FRG)  as a realization of the Wilsonian renormalization group. 
In fact, we have demonstrated that an effective theory obtained in this framework enables us to treat both transitions simultaneously on equal footing from QCD. 

In particular, the confinement/deconfinement transition in the pure gluon sector is described by the nonperturbative effective potential for the Polyakov loop average which is obtained in a nonperturbative way put forward by \cite{MP08,BGP10}
in the framework of FRG (See also \cite{BEGP10,FP13,HPS11,HSBBPSB13}). 
At present, however, the FRG studies of the Yang-Mills theory and QCD rely heavily on hard numerical works and the outcome is obtained only in the numerical way. This fact unables everyone to reproduce the FRG results and to understand the results in a physically transparent manner. Therefore, a simple analytical derivation   is desired to understand such nonperturbative results from the first principle.

We demonstrate that the essential features on the confinement/deconfinement phase transition can be obtained in a simple analytical way without hard numerical works, once we take into account a gauge-invariant and dynamical gluonic mass $M$ which is allowed to be introduced in the reformulation of the Yang-Mills theory.
In fact, we have already emphasized the importance of such a gluonic mass in the previous papers \cite{KKSS15}, but have not exhausted the outcome yet. 
The following is the summary of the investigation \cite{Kondo15}.

Besides the numerical simulations on the lattice \cite{LP13}, there are other approaches, see e.g., \cite{RSTW15,SR12,RH13,Fischer09,FK13}. 
Among them, especially, the authors of 
\cite{RSTW15} have introduced a different kind of gluonic mass term  in the gauge-fixed Yang-Mills theory at finite temperature and have investigated the effect of the mass term on confinement/deconfinement phase transition. 
They have found that the phase transition is quite well described by  the one-loop calculations in the perturbation theory, once the gluonic mass is introduced to the Yang-Mills theory. 
Their works are very interesting in its own right, but quite surprising. 
One must answer what is the meaning of the gluonic mass and 
why the one-loop calculation is so good.  
We will give a partial answer to these questions from our point of view. 
It should be remarked that their mass term is somewhat similar to ours at first glance, but its theoretical origin and the content are totally different from ours. 

\section{The strategy: our standpoint}

For this purpose, we use the reformulation of the Yang-Mills theory  \cite{KKSS15} which allows one to introduce a gauge-invariant ``mass term'' for a specific gluonic degree of freedom called the remaining field $\mathscr{X}_\mu(x)$. 
Such a gluonic mass has already played the very important role in quark confinement at zero temperature to   understand the ``Abelian dominance'' in the Maximally Abelian  gauge %
which is replaced by the gauge-independent restricted field dominance \cite{SKKMSI07} in our terminology.

The standpoint of our approach, the first approximation and its improvements, is completely different from the other work based on the systematic loop calculations in the perturbation theory \cite{RSTW15}.
The standpoint of our approach has been explained in the previous work \cite{Kondo10}. 
We aim at a purely non-perturbative approach in which we look for \textit{the initial approximation which captures the essential features of the problem in question as much as possible at the initial stage}, which is the spirit of the first approximation. 
We do not intend to do the one-loop calculation in the perturbation theory and do not intend to do the systematic loop calculations of higher orders, either.
Our approach is different from \cite{RSTW15} conceptually in this aspect.
We use the terminology ``one-loop type'' to distinguish it from the one-loop in the perturbation theory.

In the first approximation, we take into account only the quadratic terms in the fields to obtain the effective action  (except for the restricted field $\mathscr{V}_\mu(x)$), which leads to the ``one-loop type'' calculations.  It is well known that the effective action $\Gamma$ obtained from the classical action $S$ by the Legendre transform of the generating functional of the connected Green functions is equal to the classical action $S$ plus the additional part represented by the logarithmic determinant resulting from the Gaussian integrations over the quadratic parts.
Therefore, the action $S_{\rm eff}$ to be calculated by integrating out all  fields in the first approximation in our setting  agrees with the effective action $\Gamma$, up to the special treatment of the restricted field $\mathscr{V}_\mu$ as explained below.

The reason of the special treatment of the restricted field is as follows.
In our formulation, the Polyakov loop operator $L[\mathscr{A}]$ is completely written in terms of the restricted field $\mathscr{V}$, i.e., $L[\mathscr{A}]=L[\mathscr{V}]$. 
By integrating out all the fields up to the quadratic parts other than the restricted field $\mathscr{V}$, we obtain the effective theory written in terms of the  the restricted field $\mathscr{V}$ alone. 
This is along the spirit of the first approximation mentioned in the above. 
Then, we estimate the Polyakov loop average by the minimum of the effective potential obtained from the effective theory.

The resulting effective theory is identified with the low-energy effective theory in the following sense.
We use the results obtained in the first approximation  as the input for performing the FRG approach to improve the first result.  
In this sense, the first approximation is regarded as the initial condition corresponding to the large flow parameter $\kappa$ at which the FRG analysis start.
Or the first approximation can be regarded as a preliminary Ansatz for solving the flow equation of FRG. 

Of course, the above setting is just an approximation and cannot be the rigorous treatment and hence this first approximation must and will be improved afterwards by a systematic method. 
In fact, we intend to improve the first result by the non-perturbative FRG  at once (not by the systematic order by order loop expansion). 
This is the standpoint of our approach adopted in this work. 

\section{Summary of the results}

The following results are obtained based on an analytical calculation of the effective potential $V_{\rm eff}(L)$ of the Polyakov loop average $L$ alone in the $SU(2)$ and $SU(3)$ Yang-Mills theories at finite temperature $T$ in $D=4$ dimensions 
by including the gauge-invariant dynamical ``gluonic mass'' $M$.

\begin{enumerate}

\item
There exists a confinement/deconfinement phase transition at a critical temperature $T_d$ in the respective Yang-Mills theory at finite temperature $T$ signaled by the Polyakov loop average $\langle L(\bm{x}) \rangle$, i.e., 
non-vanishing $\langle L(\bm{x}) \rangle \neq 0$ for high temperature $T>T_d$, and vanishing $\langle L(\bm{x}) \rangle = 0$ for low temperature $T<T_d$.
The $Z(N)$ center symmetry which is spontaneously broken at high temperature restores at low temperature.%

\item
The critical temperature $T_d$ is estimated in the form of the ratio to the dynamical gluonic mass $M$ in the respective Yang-Mills theory:
\begin{align}
 T_d/M =& 0.34 \ \text{for  $SU(2)$} , \quad
T_d/M =  
 0.36 \ \text{for  $SU(3)$} . 
\end{align}
It should be emphasized that this ratio is gauge-independent. 
To obtain the critical temperature $T_d$, we need to know the value  $M$ of the gluonic mass.%
\footnote{
Our estimate on $T_d$ is indeed a little bit higher than expected at present. But this is based on the value of the mass $M$ obtained at zero temperature $T=0$.  The gluonic mass $M$ should depend on the temperature $T$. The mass $M$ should be determined in a self-consistent way, not just a given parameter.  
Indeed, if the mass $M$ decreases as the temperature increases: $M(T>0)<M(T=0)$, then the initial value reproduces a better result than the naive estimate.  
 Therefore, our approach has the potential to give better numerical estimate on $T_d$ without further improvements.  
The direct measurement of the gluonic mass $M$ on the lattice at finite temperature is under way.
}
The values of the gluonic mass $M$ have been measured on the lattice at zero temperature $T=0$ by 
Shibata et al. \cite{SKKMSI07}: 
\begin{align}
 M(T=0) =& 1.1  \ \text{GeV for  $SU(2)$} , \quad
 M(T=0) =  0.8 \sim 1.0 \ \text{GeV for  $SU(3)$} . 
\end{align}
A naive use of these values of $M$ leads to the estimate on $T_d$:
\begin{align}
 T_d =& 374 \ \text{MeV for  $SU(2)$} , \quad
 T_d =  288 \sim 360 \ \text{MeV for  $SU(3)$} . 
\end{align}
Incidentally, the numerical simulations on a lattice give the values \cite{LP13}:
\begin{align}
 T_d =& 295 \ \text{MeV for  $SU(2)$} , \quad
 T_d =  270   \ \text{MeV for  $SU(3)$} , 
\end{align}
while the continuum approach, e.g., the most recent FRG studies give \cite{BGP10,FP13}
\begin{align}
 T_d =& 230 \ \text{MeV for  $SU(2)$} , \quad
 T_d =  275  \ \text{MeV for  $SU(3)$} . 
\end{align}

\item
The order of the phase transition at $T_d$ is the second  order for $SU(2)$ and (weakly) first order for $SU(3)$ Yang-Mills theory.
This result is shown to be consistent with the standard  argument based on the Landau theory of phase transition using the expansion of the effective potential $V_{\rm eff}(L)$ into the power series of the Polyakov  loop average $L$ as the order parameter.
In particular, the first order transition in the $SU(3)$ Yang-Mills theory is induced by  the cubic term $L^3$ of the Polyakov loop average $L$ in the effective potential $V_{\rm eff}(L)$.

\begin{figure}[t]
\begin{center}
\includegraphics[scale=0.65]{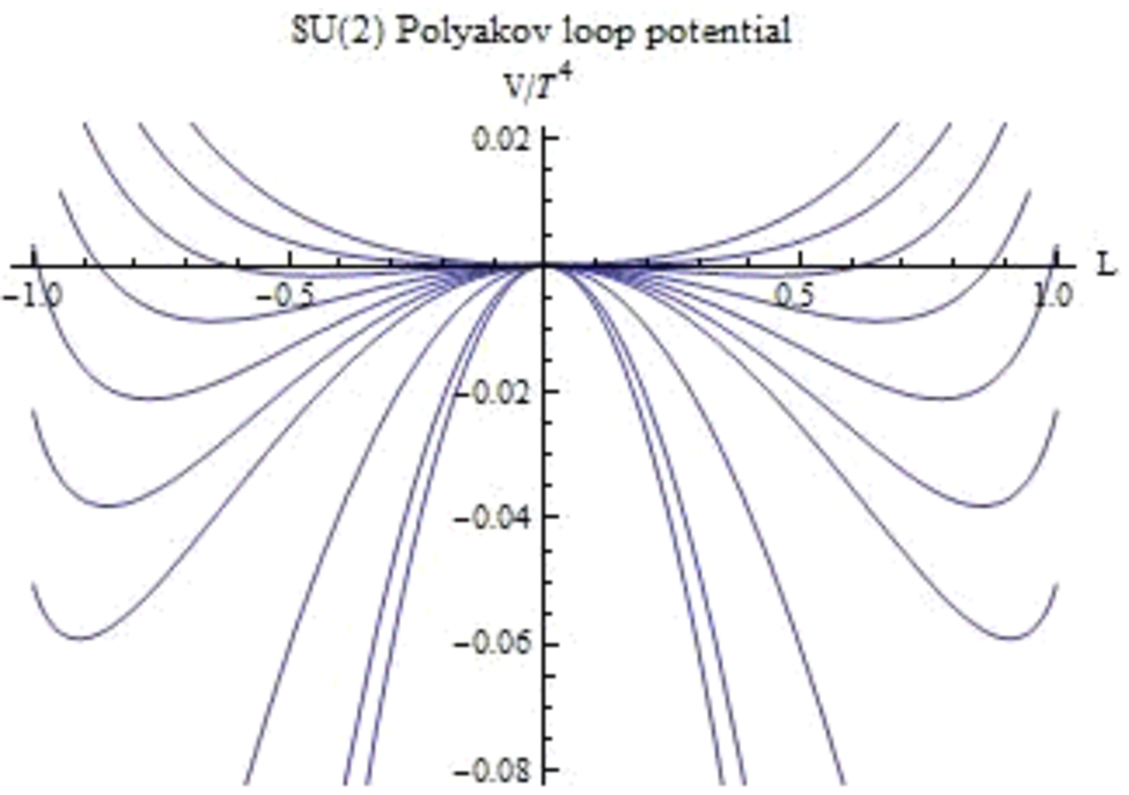}
\includegraphics[scale=0.59]{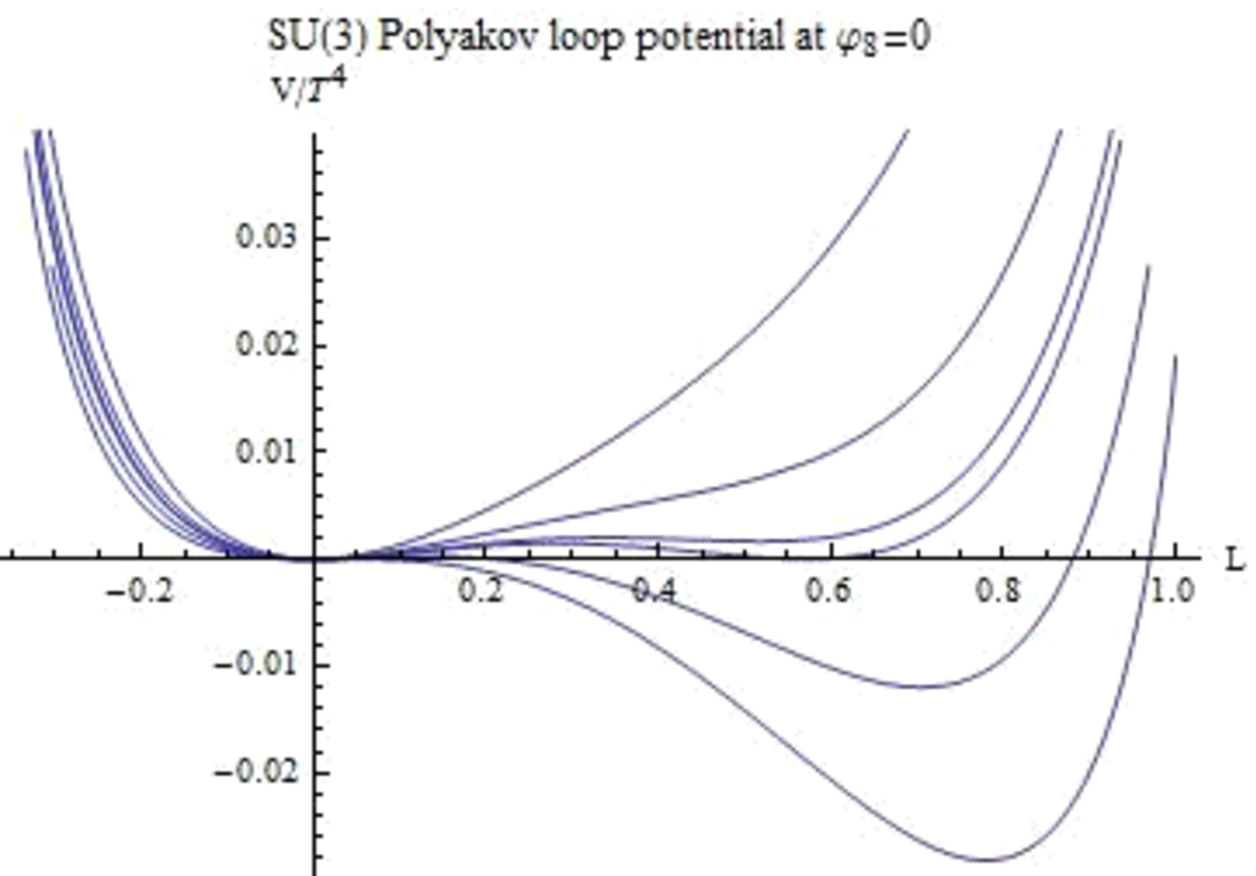}
\end{center}
\vskip -0.3cm
\caption{
(Left)
The  $D=4$ effective potential $\hat{V}$ of the $SU(2)$ Polyakov loop for   $\hat{M}:=M/T=0.0, 1.0, 2.0, 2.5, 2.6, 2.7, 2.8, 2.9, 3.0, 3.1$, 
as a function of the Polyakov loop average $L = \cos \frac{\varphi}{2} \in (-1,1]$.
(Right)
The $D=4$ effective potential $\hat{V}$ of the $SU(3)$ Polyakov loop  at $\varphi_8=0$ for $\hat{M}:=M/T=2.65, 2.70, 2.75, 2.76, 2.80, 2.90$, 
as a function of the Polyakov loop average $ L  =  \frac13 \left[ 1 + 2 \cos (\frac{\varphi_{3}}{2} )  \right] \in (-1/3,1]$, normalized as $\hat{V}(L=0)=0$.
}
\label{fig:V-SU2-all}
\end{figure}

\item
The mechanism for quark confinement or deconfinement at finite temperature is elucidated  without detailed numerical analysis in this framework by taking into account the gluonic mass $M$. 
In high temperature $T \gg M$ the gluonic mass $M$ becomes negligible and all the relevant degrees of freedom behave as massless modes, and the effective potential can be calculated in the perturbation theory 
so that the minimum of the effective potential $V_{\rm eff}(L)$ is given at the non-vanishing Polyakov loop average $L \neq 0$ implying deconfinement. 
Whereas in low temperature $T \ll M$ the ``massive'' spin-one gluonic degrees of freedom (i.e., two transverse modes and one longitudinal mode) are surpressed and the remaining unphysical massless degrees of freedom (i.e., a scalar mode, and ghost--antighost modes) become dominant. Consequently,  the signature of the effective potential $V_{\rm eff}(L)$ is reversed so that the minimum of the effective potential is given at the vanishing Polyakov loop average $L=0$ implying confinement.%
\footnote{
This observation is in line with the general arguments given in  \cite{BGP10} in the FRG and agrees with the statement given in \cite{RSTW15}.
}

\item
The above results are shown using the first approximation based on the  analytical calculations of the ``one-loop type'' (which is different from the one-loop calculation in perturbation theory).  This results of the first approximation  offer an effective starting point for the more systematic analysis of the non-perturbative studies. %
These initial results are regarded as the initial condition in solving the flow equation of the Wetterich type  and they can be improved in a systematic way in the FRG framework according to the prescription given in the previous paper \cite{Kondo10} where   
the crossover between confinement/deconfinement and chiral symmetry breaking/restoration has been analyzed from the first principle, i.e., QCD, without explicitly introducing the gluonic mass. 
But, the FRG improvement does not change the above conclusions in an essential manner. 
The above $T_d$ gives a lower bound on the true critical temperature $T_c$, since the flow evolves towards enhancing the confinement, under the assumption that $M$ does not change so much along the flow.

\end{enumerate}

\section{Remark}

We must be cautious in treating the thermodynamic observables, which needs the value of the absolute minimum $V_{\rm eff}^{\rm min}= V_{\rm eff}(L_{\rm min})$ of the effective potential $V_{\rm eff}$, i.e., the vacuum energy. We do not need such information to derive the above results which are obtained only from the location $L_{\rm min}$ of $L$ giving the minimum $V_{\rm eff}^{\rm min}$:  
\begin{align}
 V^\prime_{\rm eff}(L_{\rm min}) := \frac{\partial V_{\rm eff}(L)}{\partial L}\Big|_{L=L_{\rm min}}=0 .
\end{align}
The thermodynamic  pressure $P(T)=-V_{\rm eff}^{\rm min}(T)=- V_{\rm eff}(L_{\rm min}(T))$  remains positive in the low-temperature confined phase $L=0$ in the first approximation of our formulation, in sharp contrast to the positivity violation reported in the preceding work at one loop \cite{SR12,RSTW15}.%
\footnote{
It has been shown that the two-loop corrections improve this problem and provide both a positive pressure and a positive entropy in the whole range of temperatures explored in the second paper of \cite{RSTW15}. 
Notice that the pressure (resp. the entropy) still contains $T^4$ (resp. $T^3$) contributions at small $T$, a feature which was also signaled in \cite{RSTW15} and which is potentially also present in other approaches, in disagreement with lattice data.
} 
For the entropy density $\mathcal{S}(T) := \frac{dP(T)}{dT}$, we find the positivity violation near the critical temperature and need the improvement of the naive first approximation. 
For more details on the theoretical and physical  reasons for these artifacts, see Section IV.D and V.C   of \cite{Kondo15} and \cite{KS15}.


Notice that the mechanism of the dynamical mass generation for the gluon field has been already proposed and that the dynamical gluonic mass generation has been demonstrated to occur at zero temperature in \cite{Kondo06,Kondo14}, see also \cite{Kondo04} for the related works. 
The gauge-invariant mass $M$ for the remaining field $\mathscr{X}_\mu(x)$ can be generated 
dynamically through the gauge-invariant vacuum condensation of mass dimension two:%
\footnote{
The condensate $\Phi$ is a gauge-invariant version (which is made possible in our formulation) of the BRST-invariant vacuum condensation of mass dimension-two obtained from the on-shell BRST invariant operator of mass dimension two proposed in \cite{Kondo01}.
} 
\begin{equation}
\Phi:= \left\langle  \mathscr{X}_\rho^A  \mathscr{X}^{\rho A}  \right\rangle
=  \left\langle  2{\rm tr}[ \mathscr{X}_\rho  \mathscr{X}^{\rho }  ] \right\rangle ,
\label{dim-2-vc}
\end{equation} 
which occurs due to the quartic  self-interactions among the gluons represented by the remaining fields in the Yang-Mills theory. 
The dynamical gluonic mass $M$ is obtained  from the minimum of the effective potential $V_{\rm eff}(\Phi)$ of the vacuum condensate $\Phi$, which is also written as $V_{\rm eff}(M)$. 
Another way of understanding the mass term  is also given from the viewpoint of the gluonic Higgs field, which can be elucidated only in our formulation \cite{KKSS15,Kondo15}.

The above ideas enable us to calculate the gauge-invariant dynamical gluonic mass $M$ also at finite temperature in  our reformulation.
In fact, the temperature dependence of the dynamical mass $M(T)$ is obtained from the minimum of the effective potential $V_{\rm eff}(M)$ at finite temperature. 
At the same time, we want to calculate the Polyakov loop average $L$ to discuss the confinement/deconfinement transition.  Therefore, we need to calculate the simultaneous effective potential $V_{\rm eff}(\Phi, L)$ as a function of the two variables $M$ and $L$. 
It is confirmed by comparing \cite{Kondo14} and \cite{EGP11} that the gauge-invariant vacuum condensation of mass dimension two 
$ 
\Phi
$
can be related to the well-known gauge-invariant gluon condensation of mass dimension four, i.e., 
$
 \left\langle  \mathscr{F}_{\mu\nu}^A  \mathscr{F}^{\mu\nu A}  \right\rangle
$
responsible for the trace anomaly, which determines the non-perturbative vacuum. 
In this work, however, we treat the mass $M$ just as a constant without the temperature dependence by restricting to the effective potential $V_{\rm eff}(L)$ of the Polyakov loop average $L$ alone  for simplicity. Hence, $M$ is equal to the value at zero temperature. 
The result of the effective potential $V_{\rm eff}(\Phi, L)$ will be reported in a subsequent work to determine $\Phi$ and $L$ simultaneously.

{\it Acknowledgements}\ ---
This work is  supported by Grant-in-Aid for Scientific Research (C) 24540252 and 15K05042 from Japan Society for the Promotion of Science (JSPS).

\end{document}